\documentstyle[12pt,twoside]{article}
{\catcode `\@=11 \global\let\AddToReset=\@addtoreset}
\AddToReset{equation}{section}
\renewcommand{\theequation}{\thesection.\arabic{equation}}

\def\greaterthansquiggle{\raise.3ex\hbox{$>$\kern-.75em\lower1ex\hbox{$\sim$}}}
\def\lessthansquiggle{\raise.3ex\hbox{$<$\kern-.75em\lower1ex\hbox{$\sim$}}}
\newcommand{\beq}{\begin{equation}}
\newcommand{\eeq}{\end{equation}}
\newcommand{\beqa}{\begin{eqnarray}}
\newcommand{\eeqa}{\end{eqnarray}}
\newcommand{\beqan}{\begin{eqnarray*}}
\newcommand{\eeqan}{\end{eqnarray*}}
\newcommand{\ba}{\begin{array}}
\newcommand{\ea}{\end{array}}
\newcommand{\no}{\nonumber}

\newcommand{\ol}{\overline}
\newcommand{\ra}{\rightarrow}

\newcommand{\ve}{\varepsilon}
\newcommand{\vp}{\varphi}

\newcommand{\dg}{\dagger}
\newcommand{\wt}{\widetilde}
\newcommand{\wh}{\widehat}

\newcommand{\A}{{\cal A}}
\newcommand{\B}{{\cal B}}

\newcommand{\D}{{\cal D}}

\newcommand{\cL}{{\cal L}}
\newcommand{\M}{{\cal M}}

\newcommand{\cO}{{\cal O}}

\newcommand{\Q}{{\cal Q}}

\newcommand{\st}{\stackrel}
\newcommand{\dfrac}{\displaystyle \frac}

\newcommand{\nn}{\nonumber \\}
\newcommand{\bea}{\begin{eqnarray}}
\newcommand{\eea}{\end{eqnarray}}

\def\nz{\ifmmode {I\hskip -3pt N} \else {\hbox {$I\hskip -3pt N$}}\fi}
\def\zz{\ifmmode {Z\hskip -4.8pt Z} \else
       {\hbox {$Z\hskip -4.8pt Z$}}\fi}
\def\qz{\ifmmode {Q\hskip -5.0pt\vrule height6.0pt depth 0pt
       \hskip 6pt} \else {\hbox
       {$Q\hskip -5.0pt\vrule height6.0pt depth 0pt\hskip 6pt$}}\fi}
\def\rz{\ifmmode {I\hskip -3pt R} \else {\hbox {$I\hskip -3pt R$}}\fi}
\def\cz{\ifmmode {C\hskip -4.8pt\vrule height5.8pt\hskip 6.3pt} \else
       {\hbox {$C\hskip -4.8pt\vrule height5.8pt\hskip 6.3pt$}}\fi}

\def\au{{\setbox0=\hbox{\lower1.36775ex%
\hbox{''}\kern-.05em}\dp0=.36775ex\hskip0pt\box0}}
\def\ao{{}\kern-.10em\hbox{``}}

\normalsize
\begin{document}
\bibliographystyle{plain}
\begin{titlepage}
\begin{flushright}
CPT-99/P.3884\\
UWThPh-1999-51\\
LU TP 99/17\\
\end{flushright}
\vspace{2.5cm}
\begin{center}
{\Large \bf Chiral Perturbation Theory with \\ Virtual Photons and Leptons*}
\\[40pt]
M. Knecht$^{1}$, H. Neufeld$^{2}$, H.Rupertsberger$^{2}$, 
P. Talavera$^{3}$

\vspace{1cm}
${}^{1)}$ Centre de Physique Th\'eorique, CNRS Luminy\\
 Case 907, F-13288 Marseille Cedex 9, France\\[10pt]

${}^{2)}$ Institut f\"ur Theoretische Physik, Universit\"at
Wien\\ Boltzmanngasse 5, A-1090 Wien, Austria \\[10pt]

${}^{3)}$ Department of Theoretical Physics 2, Lund University\\
S\"olvegatan 14A, S-22362 Lund, Sweden

\vfill
{\bf Abstract} \\
\end{center}
\noindent
We construct a low-energy effective field theory which allows the
full treatment of isospin-breaking effects in semileptonic weak
interactions. To this end, we enlarge the particle spectrum of
chiral perturbation theory with virtual photons by including also the 
light leptons as dynamical degrees of freedom. Using
super-heat-kernel techniques, we determine the additional
one-loop divergences generated by the presence of virtual
leptons and give the full list of associated local counterterms.
We illustrate the use of our effective theory by applying it to
the decays $\pi \ra \ell \nu_{\ell}$ and $K \ra \ell \nu_{\ell}$.
\vfill
 
\vfill
\noindent * Work supported in part by  TMR, EC-Contract No. ERBFMRX-CT980169 
(EURODA$\Phi$NE) 

\end{titlepage}
\section{Introduction}
\label{sec: Introduction}
\renewcommand{\theequation}{\arabic{section}.\arabic{equation}}
\setcounter{equation}{0}
In the last few years, the predictions of chiral perturbation
theory \cite{Weinberg79,GL84,GL85} have been carried forward to a 
remarkable degree of
accuracy. In particular, the result for the pion--pion
scattering amplitude is now available at the two-loop level
\cite{BCEGS} in the standard case. 
Let us illustrate this theoretical progress by the
following numbers: Weinberg's calculation \cite{Weinberg66} 
of the scattering amplitude at leading order
in the low-energy expansion gives the value $a_0^0 = 0.16$ for
the isospin zero S-wave scattering length. The one-loop
calculation \cite{GL83} shifts the leading order term to $a_0^0
= 0.20$. Finally, the recent analysis to order $p^6$
\cite{BCEGS} predicts the scattering length to lie within the
range $0.206 \leq a_0^0 \leq 0.217$ (see also \cite{Girlanda}). 
An analogous calculation at
next-to-next-to-leading order has also been performed in the
framework of generalized chiral perturbation theory \cite{KMSF},
which allows also for smaller values of the quark condensate
than the standard scheme. In the generalized framework, an accurate 
determination of $a_0^0$ within the whole
range of the present experimental value, $a_0^0 = 0.26 \pm 0.05$ 
\cite{Rosselet}, 
can be interpreted in terms of the size of 
$\langle 0|\bar{q} q|0 \rangle$,  
whereas standard chiral perturbation theory, 
resting on the assumption that $\langle 0|\bar{q} q|0 \rangle$ is large, 
gives a rather precise prediction for this scattering length.
New high statistics data which are expected for the near future
may allow one to determine the nature of chiral symmetry breaking
and to decide about the validity of the standard chiral expansion
scheme. In this context, the measurement of the lifetime of
$\pi^+ \pi^-$ atoms at CERN \cite{DIRAC}, or new high statistics
$K_{\ell 4}$ experiments by the E865 and KLOE collaborations at
BNL \cite{E865} and DA$\Phi$NE \cite{DAPHNE}, respectively, are
of particular interest.  

The theoretical results mentioned above were obtained by
neglecting all isospin-breaking effects, i.e. in the limit $m_u
= m_d$, $e = 0$. However, once even two-loop effects are taken
into account, such an approach is not sufficient any more. It
has been shown explicitly \cite{KU98} that the electromagnetic
corrections to the S-wave scattering lengths are of comparable
size to the $\cO (p^6)$ strong interaction contributions.  
(See also \cite{MMS97} for a discussion of electromagnetic
effects in neutral pion scattering.)
Such an analysis requires, of course, an extension of the usual
low-energy effective theory. While isospin-breaking effects
generated by a non-vanishing quark mass difference $m_d - m_u$
are fully contained in the pure QCD sector of the effective
chiral Lagrangian, the treatment of isospin violation of electromagnetic origin
demands the inclusion of virtual photons and the appropriate local
terms up to $\cO (e^2 p^2)$. The suitable theoretical framework   
has been worked out in \cite{Urech,NR95,ELM} for the three flavour case and in
\cite{MMS97,KU98} for chiral SU(2). With these theoretical tools, it
is possible to obtain the chiral structure of the
electromagnetic contributions to $\cO (e^2 p^2)$ for all purely
mesonic matrix elements. 

The analysis of electromagnetic corrections in semileptonic
reactions requires still a further extension of chiral
perturbation theory: In this case, also the light leptons have
to be included as explicit dynamical degrees of freedom. (An
older discussion of radiative corrections in semileptonic weak
interactions, with a current algebra treatment of the hadronic matrix 
elements involved, can be found in \cite{Sirlin78}.) Only
within such a framework, one will have full control over all
possible isospin-breaking effects in the analysis of high
statistics $K_{\ell 4}$ data which will constitute an important
source of information on the $\pi - \pi$ scattering parameters and low-energy 
phase shifts.
The same refined methods are, of course, also necessary for the
interpretation of forthcoming high precision experiments on
other semileptonic decays like $K_{\ell 3}$, etc.

It is the purpose of this paper to lay the necessary theoretical
foundations for the full treatment of electromagnetic effects in
the semileptonic processes of the pseudoscalar octet within the
framework of an effective low-energy theory. In Sect. 2
we construct our lowest-order effective Lagrangian and we define
our (extended) chiral counting rules. The
additional one-loop divergences generated by the presence of
virtual leptons are obtained in Sect. 3 by using recently developed
super-heat-kernel methods \cite{Berezinian,SHK}. 
The full list of local counterterms
arising at next-to-leading order is presented in Sect.4. In
Sect. 5 we apply our effective theory to the decays $\pi \ra
\ell \nu_{\ell}$ and $K \ra \ell \nu_{\ell}$. Our conclusions,
together with an outlook to  possible applications and
extensions of the present work, are summarized in Sect. 6.
Several expressions which would only interrupt the
argument in the text are collected in the Appendix.   

\section{The effective Lagrangian to lowest order}
\label{sec: Leff} 
\renewcommand{\theequation}{\arabic{section}.\arabic{equation}}
\setcounter{equation}{0}
For a complete treatment of electromagnetic effects in the semileptonic
decays of pions and kaons, not only the pseudoscalars but also the photon
and the light leptons have to be included as dynamical degrees of freedom
in an appropriate effective Lagrangian. Its construction starts with QCD 
in the limit $m_u = m_d = m_s = 0$. The resulting symmetry
under the chiral group $G = SU(3)_L \times SU(3)_R$ is spontaneously
broken to $SU(3)_V$. The pseudoscalar mesons $(\pi,K,\eta)$ are interpreted
as the corresponding Goldstone fields $\vp_i$ ($i = 1,\ldots,8$) acting
as coordinates of the coset space $SU(3)_L \times SU(3)_R/SU(3)_V$.
The coset variables $u_{L,R}(\vp)$ are transforming as
\beqa
u_L(\vp) &\st{G}{\ra}& g_L u_L h(g,\vp)^{-1}, \no \\
u_R(\vp) &\st{G}{\ra}& g_R u_R h(g,\vp)^{-1}, \no \\
g = (g_L,g_R) &\in& SU(3)_L \times SU(3)_R,
\eeqa
where $h(g,\vp)$ is the nonlinear realization of $G$ \cite{CCWZ}.

The photon field $A_\mu$ and the leptons $\ell,\nu_\ell$ ($\ell = e,\mu$)
are introduced in
\beq
u_\mu = i [u_R^\dg (\partial_\mu - i r_\mu)u_R - u_L^\dg
(\partial_\mu - i l_\mu)u_L]
\eeq
by adding appropriate terms to the usual external vector and axial-vector
sources $v_\mu$, $a_\mu$. At the quark level, this procedure
corresponds to the usual 
minimal coupling prescription in the case of electromagnetism, and to Cabibbo 
universality in the case of the charged weak currents:
\beqa \label{sources}
l_\mu &=& v_\mu - a_\mu - e Q_L^{\rm em} A_\mu + \sum_\ell
(\bar \ell \gamma_\mu \nu_{\ell L} Q_L^{\rm w} + \ol{\nu_{\ell L}} 
\gamma_\mu \ell
Q_L^{{\rm w}\dg}), \no \\
r_\mu &=& v_\mu + a_\mu - e Q_R^{\rm em} A_\mu.
\eeqa
The $3 \times 3$ matrices $Q_{L,R}^{\rm em}$, $Q_L^{\rm w}$ are 
spurion fields with the transformation properties
\beq
Q_L^{\rm em,w} \st{G}{\ra} g_L Q_L^{\rm em,w} g_L^\dg, \qquad
Q_R^{\rm em} \st{G}{\ra} g_R Q_R^{\rm em} g_R^\dg
\eeq
under the chiral group. We also define
\beq \label{Qhom}
\Q_L^{\rm em,w} := u_L^\dg Q_L^{\rm em,w} u_L, \qquad
\Q_R^{\rm em} := u_R^\dg Q_R^{\rm em} u_R
\eeq
transforming as
\beqa
\Q_L^{\rm em,w} &\st{G}{\ra}& h(g,\vp) \Q_L^{\rm em,w} h(g,\vp)^{-1}, 
\no \\
\Q_R^{\rm em} &\st{G}{\ra}& h(g,\vp) \Q_R^{\rm em} h(g,\vp)^{-1}.
\eeqa
At the end, one identifies $Q_{L,R}^{\rm em}$ with the quark charge matrix
\beq \label{Qem}
Q^{\rm em} = \left[ \ba{ccc} 2/3 & 0 & 0 \\ 0 & -1/3 & 0 \\ 0 & 0 & -1/3 \ea
\right],
\eeq
whereas the weak spurion is taken at
\beq \label{Qw}
Q_L^{\rm w} = - 2 \sqrt{2}\; G_F \left[ \ba{ccc}
0 & V_{ud} & V_{us} \\ 0 & 0 & 0 \\ 0 & 0 & 0 \ea \right],
\eeq
where $G_F$ is the Fermi coupling constant and $V_{ud}$, $V_{us}$ are
Kobayashi--Maskawa matrix elements.

With these building blocks, our lowest order effective Lagrangian takes
the form
\beqa \label{Leff}
\cL_{\rm eff} &=& \frac{{F_0}^2}{4} \; \langle u_\mu u^\mu + \chi_+\rangle +
e^2 {F_0}^4 Z \langle \Q_L^{\rm em} \Q_R^{\rm em}\rangle \no \\
&& \mbox{} - \frac{1}{4} F_{\mu\nu} F^{\mu\nu} + \sum_\ell
[ \bar \ell (i \! \not\!\partial + e \! \not\!\!A - m_\ell)\ell +
\ol{\nu_{\ell L}} \, i \! \not\!\partial \nu_{\ell L}],
\eeqa
where $\langle \;\rangle$ denotes the trace in three-dimensional flavour
space. $F_0$ is the pion decay constant in the chiral limit and in 
the absence of electroweak interactions. Explicit chiral symmetry breaking 
is included in
$\chi_+ = u_R^\dg \chi u_L + u_L^\dg \chi^\dg u_R$ by the substitution
$\chi \ra 2{B_0} \M_{\rm quark}$, where $B_0$ is related to the
quark condensate in the chiral limit by 
$\langle 0|\bar{q} q|0 \rangle = -F_0^2 B_0$.
 
We adopt an expansion scheme where
the electric charge $e$,  the lepton masses $m_e, m_\mu$ and
fermion bilinears are considered as quantities of order $p$ in the 
chiral counting, where $p$ is a typical meson momentum.
Note, however, that terms of $\cO (e^4)$ will be neglected throughout.

\section{One-loop divergences}
For the construction of the one-loop functional we first add a gauge-breaking
term (we are using the Feynman gauge) and external sources to (\ref{Leff}):
\beq
\cL_{\rm eff} \ra \cL_{\rm eff} - \frac{1}{2}(\partial_\mu A^\mu)^2 -
J_\mu A^\mu + \sum_\ell (\bar \rho_\ell \ell + \bar \ell \rho_\ell +
\bar \sigma_\ell \nu_{\ell L} + \ol{\nu_{\ell L}} \sigma_\ell). \label{Lsource}
\eeq
Then we expand the lowest-order action associated with (\ref{Lsource}) 
around the solutions $\vp_{\rm cl}$, $A^\mu_{\rm cl}$, $\ell_{\rm cl}$,
$\nu_{\ell L,\rm cl}$ of the classical equations of motion.
In the standard ``gauge'' $u_R(\vp_{\rm cl}) = u_L(\vp_{\rm cl})^\dg =:
u(\vp_{\rm cl})$ a convenient choice of the pseudoscalar fluctuation
variables $\xi_i$ ($i = 1,\ldots,8$) is given by
\beq
u_R = u_{\rm cl} e^{i \xi_i \lambda_i/2{F_0}}, \qquad
u_L = u^{\dg}_{\rm cl} e^{-i \xi_i \lambda_i/2{F_0}}, \qquad
\xi_i(\vp_{\rm cl}) = 0,
\eeq
with the Gell-Mann matrices $\lambda_i$ ($i = 1,\ldots,8$).
For the photon and the fermions we write
\beq
A^\mu = A^\mu_{\rm cl} + \ve^\mu, \qquad
\ell = \ell_{\rm cl} + \eta_\ell, \qquad
\nu_{\ell L} = \nu_{\ell L,\rm cl} + \zeta_{\ell L}.
\eeq
In the following formulas, we shall drop the subscript ``${\rm cl}$''
for simplicity.
The classical equations of motion take the form
\beqa \label{EOM}
\nabla_\mu u^\mu &=& \frac{i}{2} \left(\chi_- -
\frac{1}{3} \langle \chi_-\rangle \right) + 
2 ie^2 {F_0}^2 Z [\Q_R^{\rm em}, \Q_L^{\rm em}], \nn
\Box A_\mu &=& J_\mu + \frac{e {F_0}^2}{2} \langle u_\mu
(\Q_R^{\rm em} - \Q_L^{\rm em})\rangle - e \sum_\ell \ol{\ell}
\gamma_\mu \ell, \nn
(i \not\!\partial + e \not\!\!A - m_\ell)\ell &=&
- \rho_\ell + \frac{{F_0}^2}{2} \langle u_\mu \Q_L^{\rm w}\rangle
\gamma^\mu \nu_{\ell L}, \nn
i \not\!\partial \nu_{\ell L} &=& - \sigma_\ell + \frac{{F_0}^2}{2}
\langle u_\mu \Q_L^{\rm w \dg}\rangle \gamma^\mu 
\ell_L, 
\eeqa
where
\beqa
\nabla_\mu &=& \partial_\mu + [\Gamma_\mu,\quad], \no \\
\Gamma_\mu &=& \frac{1}{2} [u^\dg (\partial_\mu - i r_\mu)u
+ u(\partial_\mu - i l_\mu) u^\dg], \no \\
\chi_- &=& u^\dg \chi u^\dg - u \chi^\dg u.
\eeqa
The solutions of (\ref{EOM}) are uniquely determined 
functionals of the external sources $v_\mu$, $a_\mu$, $\chi$,
$J_\mu$, $\rho_{\ell}$, $\sigma_{\ell}$. (Note that the usual Feynman 
boundary conditions are always implicitly understood.)

Expanding (\ref{Lsource}) up to terms quadratic in the variables $\xi_i$,
$\ve_\mu$, $\eta_\ell$, $\zeta_{\ell L}$, we obtain the second-order
fluctuation Lagrangian
\beqa \label{Lfluc}
\cL^{(2)} &=& - \frac{1}{2} \xi_i (d \cdot d + s)_{ij} \xi_j \no \\
&& \mbox{} + \frac{1}{2} \ve_\mu \Box \ve^\mu + \frac{e^2{F_0}^2}{4}
\langle (\Q_R^{\rm em} - \Q_L^{\rm em})^2\rangle \ve_\mu \ve^\mu \no \\
&& \mbox{} - \frac{ie{F_0}}{4} \langle [u_\mu, \Q_R^{\rm em} +
\Q_L^{\rm em}]\lambda_i\rangle \xi_i \ve^\mu \no \\
&& \mbox{} + \frac{e{F_0}}{2} \langle (\Q_R^{\rm em} - \Q_L^{\rm em})
\lambda_i\rangle (d^{\mu}_{ij}\xi_j) \ve_\mu \no \\
&& \mbox{} + \sum_\ell \{ \ol{\eta_\ell} (i \! \not\!\partial + 
e \! \not\!\!A - m_\ell) \eta_\ell + \ol{\zeta_{\ell L}} \, i \!
\not\!\partial \zeta_{\ell L} \no \\
&& \mbox{} + \frac{i{F_0}}{4} \langle [u_\mu,(\ol{\ell}
\gamma^\mu \zeta_{\ell L} + \ol{\eta_\ell} \gamma^\mu \nu_{\ell L})
\Q_L^{\rm w} + h.c.]\lambda_i \rangle \xi_i \no \\
&& \mbox{} + \frac{{F_0}}{2} \langle [(\ol{\ell} \gamma_\mu 
\zeta_{\ell L} + \ol{\eta_\ell} \gamma_\mu \nu_{\ell L})
\Q_L^{\rm w} + h.c.]\lambda_i \rangle d_{ij}^{\mu} \xi_j \no \\
&& \mbox{} + \frac{e{F_0}^2}{2} \ve_\mu \langle(\Q_R^{\rm em} - 
\Q_L^{\rm em}) [(\ol{\ell} \gamma^\mu \zeta_{\ell L}
+ \ol{\eta_\ell} \gamma^\mu \nu_{\ell L})\Q_L^{\rm w} + h.c.]
\rangle + e(\ol{\eta_\ell} \not\!\ve \ell + \bar \ell \not\!\ve 
\eta_\ell) \no \\
&& \mbox{} - \frac{{F_0}^2}{2} \langle u_\mu(\ol{\eta_\ell}
\gamma^\mu \zeta_{\ell L} \Q_L^{\rm w} + h.c.)\rangle\},
\eeqa
where
\beqa \label{gmu}
d^{\mu}_{ij} &=& \delta_{ij} \partial^{\mu}  
- \frac{1}{2} \langle \Gamma^\mu
[\lambda_i,\lambda_j]\rangle, 
\eeqa
and
\beqa \label{sij}
s_{ij} &=& \langle \frac{1}{8} (u \cdot u + \chi_+) \{\lambda_i,\lambda_j\}
- \frac{1}{4} u_\mu \lambda_i u^\mu \lambda_j \no \\
&& \mbox{} + \frac{e^2{F_0}^2Z}{2} \{\Q_R^{\rm em},\Q_L^{\rm em}\}
\{ \lambda_i,\lambda_j\} \no \\
&& \mbox{} - e^2{F_0}^2Z (\lambda_i \Q_R^{\rm em} \lambda_j \Q_L^{\rm em} +
\lambda_j \Q_R^{\rm em} \lambda_i \Q_L^{\rm em})\rangle. 
\eeqa

For the following intermediate steps it is convenient to switch to 
Euclidean space. The second-order fluctuation Lagrangian can then be
written in the form
\beqa \label{L2E}
\cL^{(2)} &=& \frac{1}{2} \Phi^T(-D \cdot D + Y)\Phi + \ol{\Psi}(\not\!\!\D + m)\Psi
\no \\
&& \mbox{} + i \ol{\Psi} \gamma_\rho(\beta D_\rho + \alpha_\rho)\Phi +
i \Phi^T(- \st{\gets}{D}_\rho \bar \beta + \bar \alpha_\rho)\gamma_\rho
\Psi.
\eeqa
In (\ref{L2E}) all bosonic fluctuation variables are collected
in a multicomponent field
\beq
\Phi = \left[ \ba{c} \xi_i \\ \ve_\mu \ea \right],
\eeq
where $D_\rho$ denotes an associated covariant derivative:
\beq
D_\rho = \partial_\rho + X_\rho .
\eeq
Analogously, the fermionic fluctuation fields are combined in
\beq
\Psi = \left[ \ba{c} \eta_a \\ \zeta_{mL} \ea \right] ,
\eeq
with a covariant derivative
\beq
\D_\rho = \partial_\rho + i f_\rho .
\eeq
The explicit expressions for the matrix-fields $X_\rho$, $Y$, $f_\rho$,
$m$, $\beta$, $\alpha_\rho$ are given in the Appendix.

In the notation of \cite{Berezinian}, the second-order
fluctuation action associated with (\ref{L2E}) has the general structure
\beq
\label{Sfluc}
S^{(2)} = \dfrac{1}{2}\Phi^T\A\Phi + \ol \Psi \B \Psi 
+ \Phi^T \ol \Gamma \Psi +
\ol \Psi \Gamma \Phi.
\eeq
The one-loop functional $W_{L = 1}$ is given by the Gaussian functional
integral
\beqa
\label{Gauss}
e^{-W_{L = 1}} =
\int [d\Phi d\Psi d\ol \Psi] ~
e^{- S^{(2)}},
\eeqa
leading to
\beq
\label{WEL1}
W_{L = 1} =  \dfrac{1}{2} \mbox{ Tr } \ln \dfrac{\A}{\A_0}
- \mbox{Tr } \ln \dfrac{\B}{\B_0} -
\sum_{n=1}^\infty \dfrac{1}{2n}\mbox{Tr }\left(\A^{-1} \ol \Gamma \B^{-1} 
\Gamma
- \A^{-1} \Gamma^T \B^{-1T} \ol \Gamma^T \right)^n ,
\eeq
where $\A_0$, $\B_0$ denote the free-field limit of $\A$ and $\B$, 
respectively.

The term $\frac{1}{2} \mbox{ Tr } \ln (\A/\A_0)$ corresponds to
purely bosonic loops (only pseudoscalar and/or photon propagators in
the loop), its divergent part has been calculated by Gasser and
Leutwyler \cite{GL85} for the strong sector and by Urech
\cite{Urech} for the electromagnetic sector. In our theory, the
divergences generated by the bosonic loops have exactly the same
form, however with the generalized source term $l_\mu$ defined
in (\ref{sources}). As a consequence, also contributions with
external lepton fields are induced in this way.

Purely fermionic loops are described by $- \mbox{Tr } \ln
(\B/\B_0)$. The divergent part of this term affects only the
wave-function renormalization of the photon, as contributions of
order $G_F^2$ will be neglected.

The last term in (\ref{WEL1}) is a genuinely new part occurring
in our effective theory with virtual leptons. It describes all one-loop graphs
where boson and fermion propagators alternate. In order to
determine the ultraviolet divergences associated with this mixed
one-loop functional, we employ the super-heat-kernel
technique developed in \cite{Berezinian}. To this end we
introduce the supermatrix
\beq
K = \left[ \ba{cc}
\A & \sqrt{2\mu}~\ol \Gamma  \\
\sqrt{2\mu} ~ \Gamma & \mu \B 
\ea \right] ,
 \label{KPP}
\eeq
where $\mu$ is an arbitrary mass parameter. The one-loop
functional can now be written as
\beq
W_{L=1} = \dfrac{1}{2} \mbox{ Str } \ln \dfrac{K}{K_0}
- \dfrac{1}{2} \mbox{ Tr } \ln \dfrac{\B}{\B_0}
 + \ldots . \label{Wss}
\eeq
The dots refer to terms at least quartic in the fermion fields
which are irrelevant for our present purposes.
In the proper-time formulation, the first term in (\ref{Wss})
assumes the form
\beqa
\dfrac{1}{2} \mbox{ Str } \ln \dfrac{K}{K_0} &=& 
- \dfrac{1}{2} \int_0^\infty \dfrac{d\tau}{\tau} \mbox{ Str } 
\left(e^{-\tau K}-e^{-\tau K_0}\right) \no \\
&=&  - \dfrac{1}{2} \int_0^\infty \dfrac{d\tau}{\tau} \int d^dx 
\mbox{ str } 
\langle x|e^{-\tau K}-e^{-\tau K_0}|x \rangle . 
\label{ptf}
\eeqa
The further evaluation of this expression is considerably
simplified \cite{SHK} by the observation that the action
associated with (\ref{L2E}) is invariant under local gauge transformations
\beq
\ba{llll}
\Phi(x) &\ra& R(x) \Phi(x), & R(x)^T R(x) =  {\bf 1},  \\
\Psi(x) &\ra& U(x) \Psi(x), & U(x)^{\dagger} U(x) = {\bf 1},  \\
X_\mu &\ra& R \partial_\mu R^{-1} + R X_\mu R^{-1}, & \\
Y &\ra& R Y R^{-1}, & \\
i f_\mu &\ra& U \partial_\mu U^{-1} + U i f_\mu U^{-1}, & \\
\alpha_\mu &\ra& U \alpha_\mu R^{-1}, & \\
\beta &\ra& U \beta R^{-1}, & \\
m &\ra& U m U^{-1}. &
\ea
\label{gaugetransf}
\eeq
Consequently, also the divergent part of the one-loop functional exhibits 
this symmetry property \cite{thooft}. The matrix-fields $Y$, $\alpha_\mu$,
$\beta$, $m$
together with their covariant derivatives
\beqa
{\nabla}_\mu Y &:=& \partial_\mu Y + [X_\mu,Y], \no \\ 
{\nabla}_\mu \alpha_\nu &:=& \partial_\mu \alpha_\nu + i f_\mu
\alpha_\nu - \alpha_\nu X_\mu, \no
\\
{\nabla}_\mu \beta &:=& \partial_\mu \beta 
+ i f_\mu \beta - \beta X_\mu, \no \\
\nabla_\mu m &:=& \partial_\mu m + i [f_\mu,m], 
\label{covdev} 
\eeqa
and the associated ``field-strength'' tensors
\beqa
X_{\mu \nu} &:=& \partial_\mu X_\nu - \partial_\nu X_\mu + [X_\mu,X_\nu], 
\no \\
f_{\mu \nu} &:=& \partial_\mu f_\nu - \partial_\nu f_\mu + i [f_\mu,f_\nu] 
\label{fieldstrengths}
\eeqa
are therefore the appropriate building blocks for the construction 
of $W_{L=1}^{\rm div}$. As a consequence, the following
intermediate steps in the further computation of (\ref{ptf}) may be performed 
with constant fields \cite{Ball} $X_\mu$, $Y = X^2$, $\alpha_\mu$, $\beta$, 
$f_\mu$, 
$m$. As the final result for the one-loop divergences has to be 
gauge-invariant, no information 
is lost and the full expression for $x$-dependent fields is recovered 
by the  substitutions
\beqa
X^2 &\ra& Y, \no \\
\left [X_\mu,X^2 \right] &\ra& {\nabla}_\mu Y, \no \\
i f_\mu \alpha_\nu - \alpha_\nu X_\mu &\ra& {\nabla}_\mu
\alpha_\nu, \no \\
i f_\mu \beta - \beta X_\mu &\ra& {\nabla}_\mu \beta, \no \\
i \left [ f_\mu, m \right ] &\ra& \nabla_\mu m, \no \\
\left [ X_\mu,X_\nu \right ] &\ra& X_{\mu \nu}, \no \\
i \left [ f_\mu,f_\nu \right ] &\ra& f_{\mu \nu}.
\label{substitutions}
\eeqa
In this approach, the relevant diagonal matrix-element in
(\ref{ptf}) assumes the simple form
\beqa
\mbox{str } \langle x|e^{-\tau K}|x \rangle &=& \mbox{str} 
\int d^dk \langle x|e^{-\tau
K}|k \rangle \langle k|x \rangle =
\mbox{str} \int \dfrac{d^dk}{(2\pi)^d}e^{-ik \cdot x} e^{-\tau K}
e^{ik \cdot x} \no \\ 
 &=&  \mbox{str} \int \dfrac{d^dk}{(2\pi)^d} e^{M+N}, 
\label{matrel}
\eeqa
with
\beqa
M &=& - \tau \left[ \ba{cc}
k^2 - 2 i k\cdot X & 0 \\
0 & \mu (i \! \! \not\!k + i \! \! \not\!f + m) \ea \right], \no \\ [4pt]
N &=& -i\tau \sqrt{2 \mu} \left[ \ba{cc}
0 & -(i k_\mu +X_\mu) \ol \beta \gamma_\mu + \ol \alpha \cdot 
\gamma  \\
\gamma_\mu \beta (i k_\mu + X_\mu) + \gamma \cdot \alpha  & 0
\ea \right] . \label{MN}
\eeqa

We are interested only in the part bilinear in the fermionic
matrix $N$. The corresponding piece of the generating functional 
(\ref{WEL1}) is just
\beq
W_{L=1}|_{\ol \Gamma \ldots \Gamma} := -  \mbox{ Tr } (\A^{-1} \ol \Gamma
\B^{-1} \Gamma).
\eeq
The appropriate decomposition of the exponential in (\ref{matrel})
can be performed by using the formula:
\bea
\label{Duhamel}
\mbox{exp}(M+N) = \mbox{exp} \, M ~ \mbox{P}\!_s \, \mbox{exp} 
\int_0^1 ds \, \wt{N} (s)                 
\eea
with
\bea
\wt{N} (s) := e^{-sM} N e^{sM} \no
\eea
and
\bea
\mbox{P}\!_s \, \mbox{exp} \int_0^1 ds \, \wt{N} (s) :=
\sum_{n=0}^{\infty} \int_0^1 ds_1 \int_0^{s_1} ds_2 \ldots \int_0^{s_{n-1}}
ds_n \, \wt{N} (s_1) \wt{N} (s_2) \ldots \wt{N} (s_n).
\eea
Picking out the part bilinear in $N$,
\beq
\mbox{ str } e^{M+N} = \int_0^1 ds \int_0^s ds'
\mbox{ str }\left[ e^{(1-s)M}N e^{(s-s')M}N e^{s'M} \right] +
\ldots, 
\label{dtheorem}
\eeq
 a few simple manipulations lead to 
\beqa
\mbox{ str }  e^{M+N} 
&=& - 2 \mu \tau^2   \sum_{n=0}^{\infty} \frac{1}{n !} \int_0^1 dz \,
\mbox{ tr } \left \{ (2 i \tau z k \cdot X)^n 
\left[-(i k_\mu +X_\mu) \ol \beta + \ol \alpha_\mu
\right] \gamma_\mu \right. \no \\
&&  \qquad \left. 
\exp \left[- \tau z k^2 - \tau \mu (1-z) (i \! \! \not\!k + i \! \!
\not\!f + m)
\right] \right. \no \\
&& \qquad \left. 
\gamma_\nu \left[ \beta (i k_\nu + X_\nu) + \alpha_\nu \right] 
\vphantom{(2 i \tau z k \cdot X)^n 
\left[-(i k_\mu +X_\mu) \ol \beta + \ol \alpha_\mu
\right] \gamma_\mu} \right\}
+ \ldots . 
\eeqa
After integration over $z$, the $\mu$-dependent terms cancel once the
proper-time and the momentum-space integrals are applied. The
remaining contribution to $W_{L=1}$ assumes the form
\beqa
W_{L=1}|_{\ol \Gamma \ldots \Gamma}
&=&  \int d^dx \sum_{n=0}^{\infty} \int_0^\infty \dfrac{dt}{t} 
t^{n+3-d} \int \dfrac{d^dl}{(2\pi)^d} \, (l^2)^{-n-1}  \mbox{ tr }
\left \{ 
\vphantom{\left[-(i l_\mu/t +X_\mu) \ol \beta + \ol \alpha_\mu
\right]}
(2 i l \cdot X)^n 
\right. \nn
&&  \qquad \left.
\left[-(i l_\mu/t +X_\mu) \ol \beta + \ol \alpha_\mu
\right] \gamma_\mu 
\exp \left(- i \! \! \not l - i t \! \! \not\!f - t m \right)
\right. \nn
&& \qquad \left.
\gamma_\nu \left[ \beta (i l_\nu/t + X_\nu) + \alpha_\nu \right]  
\vphantom{\left[-(i l_\mu/t +X_\mu) \ol \beta + \ol \alpha_\mu
\right]}        
\right \},
\eeqa
where a suitable change of the integration variables has been performed.
The divergent part (for $d \ra 4$) can now be easily isolated.
Note that terms of the form $\ol \beta \ldots \beta$ are at
least quadratic in $G_F$ (see (\ref{beta}) and (\ref{betabar}))
and will thus be discarded. For the further decomposition of 
$\exp (- i \! \! \not l - i t \! \! \not \! f - t m) $ we employ again
(\ref{Duhamel}) up to second order in $ i \! \! \not\!f + m $.
Then we perform the momentum-space integration using
\beqa
\lim_{d\to 4} \int \dfrac{d^dl}{(2\pi)^d}\left(\dfrac{\cos l}{l^2},
\dfrac{\sin l}{l^3}, \dfrac{\sin l}{l}\right) &=&
\dfrac{1}{(4\pi)^2}(-2 , 2, -4), \qquad
l:=\sqrt{l_\mu l_\mu}.
\eeqa
In the next step, one has to identify the appropriate
gauge-invariant combinations (constituting a non-trivial check
of our calculation) and reconstruct the full result by using
(\ref{substitutions}).
In this way, we obtain:
\beqa
W_{L=1}^{\rm div}|_{\ol \Gamma \ldots \Gamma} &=& 
\dfrac{1}{ {(4\pi)}^2 (d-4)} \int d^4x \mbox{ tr }
\left[
\vphantom{+ 2 \ol \alpha \cdot \gamma m^2 \beta
- 2 \ol \beta m^2 \gamma \cdot \alpha}
 \ol \alpha \cdot \gamma \gamma_\mu \nabla_\mu \gamma \cdot \alpha
- 2 \ol \alpha \cdot \gamma m \gamma \cdot \alpha 
\right. \nn
&&\left. 
\mbox{} + 2 \ol \alpha \cdot \gamma \beta Y
- 2 \ol \beta \gamma \cdot \alpha Y 
- \ol \alpha \cdot \gamma \gamma_\mu \gamma_\nu \nabla_\mu \nabla_\nu \beta 
+ \ol \beta \gamma_\mu \gamma_\nu \nabla_\mu \nabla_\nu \gamma \cdot \alpha
\right. \nn
&&\left.   
\mbox{} - \ol \alpha \cdot \gamma (\not\!\nabla m) \beta
+ \ol \alpha \cdot \gamma m \beta \gamma_\mu \nabla_\mu \beta
- 2 \ol \beta (\not\!\nabla m) \gamma \cdot \alpha
- \ol \beta m \gamma_\mu \nabla_\mu \gamma \cdot \alpha
\right. \nn 
&& \left.  
\mbox{} + 2 \ol \alpha \cdot \gamma m^2 \beta
- 2 \ol \beta m^2 \gamma \cdot \alpha  
\right ]. 
\label{result}
\eeqa
Now we insert the explicit expressions for $X_\mu$, $Y$, $f_\mu$, $\alpha_\mu$,
$\beta$, $m$. Discarding again all terms which are $\cO(G_F^n)$ with 
$n \geq 2$, we get
\beqa
W_{L=1}^{\rm div}|_{\ol \Gamma \ldots \Gamma} &=& 
 - \dfrac{e^2}{ {(4\pi)}^2 (d-4)} \int d^4x \sum_{\ell}
\left \{  2 \bar \ell ( \! \not\!\partial - i e \! \not\!\!A )\ell 
        + 8 m_\ell \ol \ell  \ell \right. \nn
&& \left.
\mbox{} + \frac{3 F_0^2}{2} 
\langle (\ol{\ell} \gamma_\mu \nu_{\ell L} \Q_L^{\rm w} 
-\ol{\nu_{\ell L}} \gamma_\mu \ell \Q_L^{{\rm w}\dg})
\nabla_\mu (\Q_R^{\rm em} - \Q_L^{\rm em}) \rangle
\right. \nn
&& \left.
\mbox{} + i {F_0}^2 
\langle (\ol{\ell} \gamma_\mu \nu_{\ell L} \Q_L^{\rm w} 
- \ol{\nu_{\ell L}} \gamma_\mu \ell \Q_L^{{\rm w}\dg})
[u_\mu , \Q_L^{\rm em}] \rangle
\right. \nn
&& \left.
\mbox{} + 3 {F_0}^2  m_\ell 
\langle (\ol{\ell} \nu_{\ell L} \Q_L^{\rm w} 
+ \ol{\nu_{\ell L}} \ell \Q_L^{{\rm w}\dg})
\Q_R^{\rm em} \rangle
\right\}.
\label{diveuclid}
\eeqa

Transforming back to Minkowski space, our result for the
divergent part of the mixed one-loop functional takes the final form 
\beqa
W_{L=1}^{\rm div}|_{\ol \Gamma \ldots \Gamma} &=& 
\dfrac{e^2}{ {(4\pi)}^2 (d-4)} \int d^4x \sum_{\ell}
\left \{- 2 \bar \ell (i \! \not\!\partial + e \! \not\!\!A )\ell 
       +  8 m_\ell \ol \ell  \ell \right. \nn
&& \left.
\mbox{} -\frac{3 i {F_0}^2}{2} 
\langle (\ol{\ell} \gamma_\mu \nu_{\ell L} \Q_L^{\rm w} 
-\ol{\nu_{\ell L}} \gamma_\mu \ell \Q_L^{{\rm w}\dg})
\nabla^\mu (\Q_R^{\rm em} - \Q_L^{\rm em}) \rangle
\right. \nn
&& \left.
\mbox{} + {F_0}^2 
\langle (\ol{\ell} \gamma_\mu \nu_{\ell L} \Q_L^{\rm w} 
- \ol{\nu_{\ell L}} \gamma_\mu \ell \Q_L^{{\rm w}\dg})
[u^\mu , \Q_L^{\rm em}] \rangle
\right. \nn
&& \left.
\mbox{} + 3 {F_0}^2  m_\ell 
\langle (\ol{\ell} \nu_{\ell L} \Q_L^{\rm w} 
+ \ol{\nu_{\ell L}} \ell \Q_L^{{\rm w}\dg})
\Q_R^{\rm em} \rangle
\right\}.
\label{divMinkowski}
\eeqa

\section{The next-to-leading order Lagrangian}
\label{sec: NLO}
\renewcommand{\theequation}{\arabic{section}.\arabic{equation}}
\setcounter{equation}{0}
We are now in the position to construct the most general local
action at next-to-leading order which will also renormalize the 
one-loop divergences discussed in the previous section.

In the absence of virtual photons and leptons, the local action of 
$\cO(p^4)$ is generated by the well-known Gasser--Leutwyler Lagrangian 
$\cL_{p^4}$
\cite{GL85}:
\beqa \label{Lp4}
\cL_{p^4} &=& L_1 \; \langle u_\mu u^\mu \rangle^2 + L_2 \; \langle u_\mu
u^\nu\rangle \; \langle u^\mu u_\nu\rangle \no \\
&& \mbox{} + L_3 \; \langle u_\mu u^\mu u_\nu u^\nu\rangle + 
L_4 \; \langle u_\mu u^\mu\rangle \; \langle \chi_+\rangle \no \\
&& \mbox{} + L_5 \; \langle u_\mu u^\mu \chi_+\rangle + 
L_6 \; \langle \chi_+\rangle^2 + L_7 \; \langle \chi_-\rangle^2 \no \\
&& \mbox{} + \frac{1}{4} (2L_8 + L_{12}) \langle \chi_+^2\rangle + 
\frac{1}{4} (2L_8 - L_{12}) \langle \chi_-^2\rangle \no \\
&& \mbox{} - iL_9 \; \langle f_+^{\mu\nu} u_\mu u_\nu\rangle + 
\frac{1}{4} (L_{10} + 2L_{11}) \langle f_{+\mu\nu} f_+^{\mu\nu}\rangle
\no \\
&& \mbox{} - \frac{1}{4} (L_{10} - 2L_{11}) 
\langle f_{-\mu\nu} f_-^{\mu\nu}\rangle, \label{L4}
\eeqa
with
\beqa
f_{\pm}^{\mu \nu} &=& u F_L^{\mu \nu} u^{\dg} \pm 
                       u^{\dg} F_R^{\mu \nu} u, \nn
F_L^{\mu \nu} &=& \partial^{\mu} l^{\nu} - \partial^{\nu} l^{\mu}
                  - i [l^\mu,l^\nu], \nn
F_R^{\mu \nu} &=& \partial^{\mu} r^{\nu} - \partial^{\nu} r^{\mu}
                  - i [r^\mu,r^\nu].
\eeqa

If also the photon is treated as a dynamical degree of freedom,
the following local counterterms of $\cO(e^2 p^2)$ have to be added
\cite{Urech}:
\beqa
\cL_{e^2p^2} &=&
e^2 {F_0}^2 \left\{ \frac{1}{2} K_1 \; \langle (\Q^{\rm em}_L)^2 +
(\Q^{\rm em}_R)^2\rangle \; \langle u_\mu  
u^\mu\rangle \right. \no \\
&& \mbox{} + K_2 \; \langle \Q^{\rm em}_L \Q^{\rm em}_R\rangle 
\; \langle u_\mu u^\mu
\rangle \no \\
&& \mbox{} - K_3 \; [\langle \Q^{\rm em}_L u_\mu\rangle 
\; \langle \Q^{\rm em}_L u^\mu
\rangle + \langle \Q^{\rm em}_R u_\mu\rangle 
\; \langle \Q^{\rm em}_R u^\mu\rangle ] \no \\
&& \mbox{} + K_4 \; \langle \Q^{\rm em}_L u_\mu\rangle 
\; \langle \Q^{\rm em}_R u^\mu \rangle
\no \\
&& \mbox{} + K_5 \; \langle[(\Q^{\rm em}_L)^2 + (\Q^{\rm em}_R)^2] 
u_\mu u^\mu\rangle \no \\
&& \mbox{} + K_6 \; \langle (\Q^{\rm em}_L \Q^{\rm em}_R + 
\Q^{\rm em}_R \Q^{\rm em}_L) u_\mu u^\mu\rangle \no \\
&& \mbox{} + \frac{1}{2} K_7 \; \langle (\Q^{\rm em}_L)^2 
+ (\Q^{\rm em}_R)^2\rangle \; \langle \chi_+\rangle
\no \\
&& \mbox{} + K_8\; \langle \Q^{\rm em}_L \Q^{\rm em}_R\rangle 
\; \langle \chi_+\rangle \no \\
&& \mbox{}+ K_9 \; \langle [(\Q^{\rm em}_L)^2 + (\Q^{\rm em}_R)^2] 
\chi_+\rangle \no \\
&& \mbox{} + K_{10}\; \langle(\Q^{\rm em}_L \Q^{\rm em}_R 
+ \Q^{\rm em}_R \Q^{\rm em}_L) \chi_+\rangle \no \\
&& \mbox{} - K_{11} \; \langle(\Q^{\rm em}_L \Q^{\rm em}_R 
- \Q^{\rm em}_R \Q^{\rm em}_L) \chi_-\rangle \no \\
&& \mbox{}- iK_{12}\; \langle[(\wh \nabla_\mu \Q^{\rm em}_L) \Q^{\rm em}_L -
\Q^{\rm em}_L \wh \nabla_\mu \Q^{\rm em}_L 
- (\wh \nabla_\mu \Q^{\rm em}_R) \Q^{\rm em}_R + 
\Q^{\rm em}_R \wh \nabla_\mu \Q^{\rm em}_R] u^\mu\rangle \no \\
&& \mbox{}+ K_{13} \; \langle (\wh \nabla_\mu \Q^{\rm em}_L) 
(\wh \nabla^\mu \Q^{\rm em}_R)
\rangle \no \\
&& \left. \mbox{} + K_{14} \; \langle (\wh \nabla_\mu \Q^{\rm em}_L) 
(\wh \nabla^\mu \Q^{\rm em}_L) +
(\wh \nabla_\mu \Q^{\rm em}_R) (\wh \nabla^\mu \Q^{\rm em}_R)\rangle  
\vphantom{\frac{1}{2} K_1}
\right\}, 
\label{LE2P2}
\eeqa
where
\beqa \label{covder1}
\wh \nabla_\mu \Q^{\rm em}_L &=& \nabla_\mu \Q^{\rm em}_L 
+ \frac{i}{2} [u_\mu,\Q^{\rm em}_L] =
u (D_\mu Q_L^{\rm em}) u^\dg , \no \\
\wh \nabla_\mu \Q^{\rm em}_R &=& \nabla_\mu \Q^{\rm em}_R 
- \frac{i}{2} [u_\mu,\Q^{\rm em}_R] =
u^\dg (D_\mu Q^{\rm em}_R) u ,
\eeqa
with
\beqa \label{covder2}
D_\mu Q^{\rm em}_L &=& \partial_\mu Q^{\rm em}_L 
- i[l_\mu,Q^{\rm em}_L], \nn
D_\mu Q^{\rm em}_R &=& \partial_\mu Q^{\rm em}_R - i[r_\mu,Q^{\rm em}_R].
\eeqa

In the presence of virtual leptons, the structure of the
Lagrangians (\ref{L4}) and (\ref{LE2P2}) remains unchanged. The
only necessary modification is the inclusion  of the lepton term
in $l_\mu$ (see (\ref{sources})). In addition to $\cL_{p^4}$ and 
$\cL_{e^2p^2}$, we need the ``leptonic'' Lagrangian
\beqa \label{Llept}
\cL_{\rm lept} &=& 
e^2 \sum_{\ell} \left \{ {F_0}^2 \left[  
X_1 \ol{\ell} \gamma_\mu \nu_{\ell L} 
\langle u^\mu  \{ \Q_R^{\rm em}, \Q_L^{\rm w} \} \rangle 
\right. \right. \nn [-5pt]
&& \qquad \qquad \left. \left.
+ X_2 \ol{\ell} \gamma_\mu \nu_{\ell L} 
\langle u^\mu  [\Q_R^{\rm em}, \Q_L^{\rm w}] \rangle
\right. \right. \nn
&& \qquad \qquad \left. \left.
+ X_3 m_\ell \ol{\ell} \nu_{\ell L} \langle \Q_L^{\rm w} \Q_R^{\rm em} \rangle
\right. \right. \nn
&& \qquad \qquad \left. \left.
+ i X_4 \ol{\ell} \gamma_\mu \nu_{\ell L} 
\langle \Q_L^{\rm w} \wh \nabla^\mu  \Q_L^{\rm em} \rangle
\right. \right. \nn
&& \qquad \qquad \left. \left.
+ i X_5 \ol{\ell} \gamma_\mu \nu_{\ell L} 
\langle \Q_L^{\rm w} \wh \nabla^\mu  \Q_R^{\rm em} \rangle 
+ h.c. \right]  \right. \nn
&& \qquad  \qquad \left.
+ X_6 \bar \ell (i \! \not\!\partial + e \! \not\!\!A )\ell
\right. \nn
&& \qquad  \qquad \left.
+ X_7 m_\ell \ol \ell  \ell \right \}. 
\eeqa
In $\cL_{\rm lept}$  we consider only terms quadratic in the lepton
fields and at most linear in $G_F$. The coupling constants
$X_1,\ldots,X_5$ are real in the limit of CP invariance and the 
reality of $X_6$ and $X_7$ is required by the hermiticity of the
associated action. The terms with $X_{4,5}$ will not appear in
realistic physical processes as the generated amplitudes
contain an external (axial-) vector source (see (\ref{covder1}) 
and (\ref{covder2})). 

In deriving a minimal set of terms in (\ref{Llept}), we have
used partial integration, the equations of motion (\ref{EOM})
and the following relations derived from (\ref{Qhom}), (\ref{Qem}) 
and (\ref{Qw}):
\beq
\Q_L^{\rm em} \Q_L^{\rm w} = \frac{2}{3} \Q_L^{\rm w} , \qquad  
\Q_L^{\rm w} \Q_L^{\rm em}  = -\frac{1}{3} \Q_L^{\rm w} , \qquad 
\langle \Q_L^{\rm w}  \rangle = 0.
\eeq
We give here some typical examples of terms which can be
eliminated in this way:
\beqa
&&
e^2 {F_0}^2 
\ol{\ell} \gamma_\mu \nu_{\ell L} 
\langle u^\mu  [ \Q_L^{\rm em}, \Q_L^{\rm w} ] \rangle 
=
e^2 {F_0}^2   
\ol{\ell} \gamma_\mu \nu_{\ell L} 
\langle u^\mu  \Q_L^{\rm w}  \rangle
\ra
2 e^2  \bar \ell (i \! \not\!\partial + e \! \not\!\!A 
-m_{\ell})\ell, \no \\
&&
e^2 \ol{\nu_{\ell L}} \, i \! \not\!\partial \nu_{\ell L}
\ra
e^2  \bar \ell (i \! \not\!\partial + e \! \not\!\!A 
-m_{\ell})\ell .
\eeqa 

Finally, also a photon Lagrangian
\beq
\cL_{\gamma} = e^2 X_8 F_{\mu \nu} F^{\mu \nu}, \qquad 
F_{\mu \nu} = \partial_{\mu} A_{\nu} - \partial_{\nu} A_{\mu},
\eeq
has to be added. This term cancels the divergences of the photon
two-point function generated by the lepton loops. (The
loop contributions of the charged pseudoscalars are renormalized
by the $L_{10,11}$ terms in (\ref{Lp4}).)

The low--energy couplings $L_i$, $K_i$, $X_i$ arising here are
divergent (except $L_3$, $L_7$, $K_7$, $K_{13}$, $K_{14}$ and $X_1$). 
They absorb the divergences of the
one-loop graphs via the renormalization
\beqa
L_i &=& L_i^r(\mu) + \Gamma_i \Lambda(\mu) , \quad i=1,\ldots,12 , \no \\
K_i &=& K_i^r(\mu) + \Sigma_i \Lambda(\mu) , \quad i=1,\ldots,14 , \no \\
X_i &=& X_i^r(\mu) + \Xi_i \Lambda(\mu) , \quad i=1,\ldots,8 , \no \\
\Lambda(\mu) &=& \frac{\mu^{d-4}}{(4\pi)^2} \left\{ \frac{1}{d-4} -
\frac{1}{2} [\ln (4\pi) + \Gamma'(1) + 1]\right\}, \label{renorm}
\eeqa
in the dimensional regularization scheme. The coefficients
$\Gamma_i$ and $\Sigma_i$ can be found in \cite{GL85} and in
\cite{Urech}, respectively. Their values are not modified by the
presence of virtual leptons as long as contributions of
$\cO(G_F^2)$ are neglected. 

The ``new'' coefficients $\Xi_1, \ldots \Xi_7$  are 
determined by (\ref{divMinkowski}) and $\Xi_8$ is derived from
the divergent part of the lepton loops:
\beqa
&& \Xi_1 = 0, \quad \Xi_2 = -\dfrac{3}{4}, \quad 
\Xi_3 = -3, \quad \Xi_4 = -\dfrac{3}{2}, 
 \nn [5pt]   
&& \Xi_5 = \dfrac{3}{2}, \quad \Xi_6 = -5, 
\quad \Xi_7 = -1, \quad \Xi_8 = -\dfrac{4}{3}.
\eeqa

\section{$\pi \ra \ell \nu_{\ell}$ and 
$K \ra \ell \nu_{\ell}$}
\label{sec: decays}
\renewcommand{\theequation}{\arabic{section}.\arabic{equation}}
\setcounter{equation}{0}
We are now in the position to perform a complete one-loop
analysis of semileptonic pion and kaon decays including the
electromagnetic contributions of $\cO(e^2 p^2)$. As an
illustration we give here the theoretical results for the decay
rates of $\pi \ra \ell \nu_{\ell}$ and $K \ra \ell \nu_{\ell}$.
The former reaction will also serve as the reference process for
our further investigations of semileptonic decays.

The contributions of graphs without virtual leptons to the $P
\ra \ell \nu_{\ell}$ amplitudes ($P = \pi, K$) have already
been given in Eqs. (5.3) and (5.4) of \cite{ELM}.
Now we have to add also those diagrams where a virtual photon is
attached to a charged lepton line, the  contributions generated
by the counter-terms in (\ref{Llept}), and the leptonic
wave-function and mass renormalization. The decay amplitude
obtained in this way is, of course, UV-finite but still
IR-divergent. To arrive at an infrared-finite result, we
consider the observable $\Gamma(P \ra \ell \nu_{\ell} (\gamma))
:= \Gamma(P \ra \ell \nu_{\ell}) 
+ \Gamma(P \ra \ell \nu_{\ell} \gamma)$. The relevant expression
for the total probability of inner bremsstrahlung can be found in Eq. (4) of
\cite{Kinoshita}. (We have checked this formula.) 

Combining the various contributions we obtain our final result for the
$\pi_{\ell 2 (\gamma)}$ decay:  
\beqa
\Gamma(\pi \ra \ell \nu_{\ell} (\gamma)) 
&=& \frac{G_F^2 |V_{ud}|^2 {F_0}^2 m_{\ell}^2
M_{\pi^{\pm}}}{4 \pi} (1-z_{\pi \ell})^2 
\no \\
&&  \left\{ \vphantom{+  e^2 E^r(\mu) 
+ \frac{e^2}{(4\pi)^2} \left[ 3 \ln \frac{M^2_\pi}{\mu^2} + H(z_{\pi \ell})
 \right] }
1 + \frac{8}{{F_0}^2} \left[L_4^r(\mu) (M^2_\pi + 2M_K^2) + 
L_5^r (\mu) M^2_\pi\right] \right. \no \\
&& \mbox{} - \frac{1}{2(4\pi)^2 {F_0}^2} \left[ 2M^2_{\pi^\pm} \ln 
\frac{M^2_{\pi^\pm}}{\mu^2} + 2 M^2_{\pi^0} \ln \frac{M^2_{\pi^0}}{\mu^2}
\right. \no \\
&& \left. \mbox{}
+ M^2_{K^\pm} \ln \frac{M^2_{K^\pm}}{\mu^2} + M^2_{K^0} \ln
\frac{M^2_{K^0}}{\mu^2} \right] \no \\
&& \left. \mbox{} +  e^2 E^r(\mu) 
+ \frac{e^2}{(4\pi)^2} \left[ 3 \ln \frac{M^2_\pi}{\mu^2} + H(z_{\pi \ell})
 \right] \right\}, \label{piondecay}
\eeqa
with $z_{\pi \ell} = m_{\ell}^2 / M_{\pi^{\pm}}^2$, where
$m_{\ell}$ denotes now the physical mass of the charged lepton.
The parameter $E^r(\mu)$ denotes the finite (renormalized) part of the 
linear combination of coupling constants
\beq \label{E}
E := \frac{8}{3} K_1 + \frac{8}{3} K_2
           + \frac{20}{9} K_5 + \frac{20}{9} K_6
           + 4 K_{12} - \frac{4}{3} X_1 - 4 X_2 + 4 X_3 - X_6.
\eeq
The kinematical function $H(z)$ is given by
\beqa \label{kinfunc}
H(z) &=& \frac{23}{2} - \frac{3}{1-z} +11 \ln z - \frac{2 \ln
z}{1-z} - \frac{3 \ln z}{(1-z)^2} \no \\
&& \mbox{} - 8 \ln (1-z) -  \frac{4(1+z)}{1-z} \ln z \ln (1-z)  
+  \frac{8(1+z)}{1-z} \int_0^{1-z} \! \! dt \, \frac{\ln (1-t)}{t}.
\eeqa

The scale independence of (\ref{piondecay}) can be checked by using
the pertinent $\beta$ function of the coupling (\ref{E}) and the
(lowest-order) expressions for the pseudoscalar masses
\beqa
 M^2_{\pi^\pm} &=& 2{B_0} \wh m + 2 e^2 Z {F_0}^2, \no \\
 M^2_{\pi^0} &=&  2{B_0} \wh m , \no \\
 M^2_{K^\pm} &=& {B_0}\left[ (m_s + \wh m) - \frac{2\ve}{\sqrt{3}}
(m_s - \wh m)\right] + 2e^2 Z {F_0}^2, \no \\
 M^2_{\stackrel{(-)}{K}{}^0} &=& {B_0} \left[(m_s + \wh m) + 
\frac{2\ve}{\sqrt{3}} (m_s - \wh m)\right] , \no \\
 M^2_\eta &=& \frac{4}{3} {B_0}\left( m_s + \frac{\wh m}{2}\right).
\label{treemass}
\eeqa
The mixing angle $\ve$ is given by
\beq
\ve = \frac{\sqrt{3}}{4} \; \frac{m_d - m_u}{m_s - \wh m} \label{epsilon},
\eeq 
the symbol $\wh m$ stands for the mean value of the light quark masses,
\beq
\wh m = \frac{1}{2} (m_u + m_d),
\eeq
and ${B_0}$ is the vacuum condensate parameter contained in $\chi_+$.
Finally, $M_{\pi}$ and $M_K$ in (\ref{piondecay}) denote the
isospin limits ($m_u = m_d$, $e = 0$) of the pion mass and the
kaon mass, respectively:
\beq
 M^2_{\pi} =  2{B_0} \wh m , \quad 
 M^2_K = {B_0} (m_s + \wh m).
\eeq

Equation (\ref{piondecay}) allows in principle to extract the value 
of the pion decay constant $F_\pi$ from the experimental knowledge of the 
$\pi_{\ell 2 (\gamma)}$ decay rate. In order to obtain an unambiguous answer, 
let us define $F_\pi$ as being given by the matrix element of the appropriate 
axial current between the vacuum and the charged pion state in pure QCD. 
At one loop, this gives the following (scale 
independent) expression \cite{GL85}
\beqa
F_\pi &=& F_0 \bigg\{1 + \frac{4}{{F_0}^2} \left[L_4^r(\mu) (M^2_\pi + 2M_K^2)
 + 
L_5^r (\mu) M^2_\pi\right]  \no \\
&& \mbox{} - \frac{1}{2(4\pi)^2 {F_0}^2} \left[ 2 M^2_{\pi} \ln 
\frac{M^2_{\pi}}{\mu^2}  
+ M^2_{K} \ln \frac{M^2_{K}}{\mu^2}\right] \bigg\}.\label{Fpi1loop}
\eeqa

Extensive studies of the $\cO(e^2)$  radiative
corrections to $\pi_{\ell 2}$ decays already exist 
in the literature \cite{Kinoshita,elmc,Marciano/Sirlin}. Marciano and Sirlin 
\cite{Marciano/Sirlin} have summarized the presently known short- and 
long-distance
radiative corrections to $\Gamma(\pi \ra \ell \nu_{\ell} (\gamma))$.
Their result, given in Eq. (7a) of \cite{Marciano/Sirlin}, agrees
with the general structure of (\ref{piondecay}).
Note that the kinematical function $F(x)$ defined in Eq. (7b) of
\cite{Marciano/Sirlin} is related to (\ref{kinfunc}) by $F(x) =
H(x^2)/4 - 1/2$, and that the terms associated with the parameters $C_2$, 
$C_3$ of \cite{Marciano/Sirlin} are beyond the scope of our
next-to-leading order analysis. In order to make contact with 
\cite{Marciano/Sirlin}, we need to specify the physical meaning we ascribe to 
the constant $G_F$, which so far has mainly served as a book-keeping device. 
From the point of view of the low-energy approach we are following, it seems 
natural to identify $G_F$ with the muon decay constant. To lowest order 
in the weak interactions, the muon decay amplitude follows from the 
effective Lagrangian
\beq
\cL_{\rm lept}' = -\frac{G_F}{\sqrt{2}}\,
\big[\,{\bar{\mu}}\gamma_\mu(1-\gamma_5)\nu_{\mu}\,\big]\,
\big[\,{\bar{e}}\gamma^\mu(1-\gamma_5)\nu_{e}\,\big].
\eeq
When radiative corrections are included, the total muon decay rate to leading 
order in $G_F$ remains finite 
order by order in powers of the fine structure constant $\alpha$
\cite{BermanSirlin}. At order $\alpha$ \cite{KS59},
\beq
\Gamma(\mu\to {\mbox{all}}) = \frac{G_F^2m_\mu^5}{192\pi^3}\,
f\left(\frac{m_e^2}{m_\mu^2}\right)\,\big[\,1\,+\,\frac{\alpha}{2\pi}
(\frac{25}{4}-\pi^2)\,+\,\cO(\alpha^2)\,\big],
\eeq
with $f(x) = -8x -12x^2\ln x +8x^3 - x^4$ for massless neutrinos. The 
corrections of order $\alpha^2$ have recently been computed in 
\cite{RitStuart} (see also the discussion in \cite{Sirlin99}). Up to a 
negligible correction factor $1+3m_\mu^2/10M_W^2$, $G_F$ can be identified 
with the constant $G_\mu$ appearing in Eq. (7a) of \cite{Marciano/Sirlin}. If 
we further identify $f_\pi/\sqrt{2}$ occurring in the same expression with 
the QCD quantity $F_\pi$ as defined above, we obtain for the
constant $C_1$ of \cite{Marciano/Sirlin}:
\beq
C_1 = - 4\pi^2 E^r(M_{\rho}) - \frac{1}{2} + \ln\frac{M_Z^2}{M_\rho^2}
+ \frac{Z}{4}\left[ 3 + 
2\ln\frac{M_\pi^2}{M_\rho^2} + \ln\frac{M_K^2}{M_\rho^2}
\right].
\eeq

Quite analogously, we find for the $K_{\ell 2 (\gamma)}$ decay:
\beqa
\Gamma(K \ra \ell \nu_{\ell} (\gamma)) &=& 
\frac{G_{F}^2 |V_{us}|^2 {F_0}^2 m_{\ell}^2
M_{K^{\pm}}}{4 \pi} (1-z_{K \ell})^2 
\no \\
&&  \left\{ \vphantom{\frac{e^2}{2(4\pi)^2} \left[ 3 \ln \frac{M^2_K}{\mu^2} +
H(z_{K \ell}) \right] }
1 + \frac{8}{{F_0}^2} \left[ L_4^r(\mu) (M^2_\pi + 2M_K^2) + 
L_5^r (\mu) M^2_K \right] \right. \no \\
&& \mbox{} - \frac{1}{4(4\pi)^2 {F_0}^2} \left[ 2M^2_{\pi^\pm} \ln 
\frac{M^2_{\pi^\pm}}{\mu^2} +  M^2_{\pi^0} \ln \frac{M^2_{\pi^0}}{\mu^2}
\right. \no \\
&& \left.  \mbox{}
+ 4M^2_{K^\pm} \ln \frac{M^2_{K^\pm}}{\mu^2} + 2M^2_{K^0} \ln
\frac{M^2_{K^0}}{\mu^2} + 3M^2_\eta \ln \frac{M^2_\eta}{\mu^2} \right] \no \\
&&\mbox{} + \frac{16\sqrt{3}\;\ve}{3{F_0}^2} \; L_5^r(\mu) (M^2_\pi
- M^2_K) - \frac{\sqrt{3}\;\ve}{2(4\pi)^2{F_0}^2} 
\left[ M^2_\pi \ln \frac{M^2_\pi}{\mu^2} - M^2_\eta \ln
\frac{M^2_\eta}{\mu^2} \right] \no \\
&& \mbox{} + \left. e^2 E^r(\mu) 
+ \frac{e^2}{(4\pi)^2} \left[ 3 \ln \frac{M^2_K}{\mu^2} +
H(z_{K \ell}) \right] \right\}, \label{Kdecay}
\eeqa
with $z_{K \ell} = m_{\ell}^2 / M_{K^{\pm}}^2$. Note that
(\ref{piondecay}) as well as (\ref{Kdecay}) contain the same
(unknown)
electromagnetic low-energy coupling constant $E^r(\mu)$.
Taking the ratio of (\ref{Kdecay}) and (\ref{piondecay}),
$E^r(\mu)$ cancels and therefore the electromagnetic corrections to this
specific combination of observables are uniquely determined at the order 
we consider. In
principle, this result could be used for an improvement in extracting 
the (strong) low-energy coupling $L_5$ from the experimental
data. In practice, however, the uncertainties due to higher
order strong interaction effects are much bigger than the electromagnetic 
corrections. Nevertheless, this example shows that certain
observables allow for unambiguous predictions of the
associated electromagnetic contributions in spite of our
(presently) poor knowledge about the values of the electromagnetic 
low-energy couplings. The same is, of course, also true for the
ratios $\Gamma(P \ra e \nu_e (\gamma)) / 
\Gamma(P \ra \mu \nu_{\mu} (\gamma))$. 

Defining the decay constant of the charged kaon as described above for the 
pion\footnote{In the case of the pion, the distinction between the charged 
and the neutral decay constant is a tiny $\cO\big((m_d-m_u)^2\big)$ effect, 
and arises only at higher orders, whereas 
$F_{K^{\pm}}/F_{\stackrel{(-)}{K}{}^0}$ is $\cO\big((m_d-m_u)\big)$.} we
obtain \cite{GL85}
\beqa
F_{K^{\pm}} &=& F_0 \bigg\{1 + 
\frac{4}{{F_0}^2} \left[L_4^r(\mu) (M^2_\pi + 2M_K^2)
 + 
L_5^r (\mu) M^2_K\right]  \no \\
&& \mbox{} - \frac{1}{8(4\pi)^2 {F_0}^2} \left[ 3 M^2_{\pi} \ln 
\frac{M^2_{\pi}}{\mu^2}  
+ 6 M^2_{K} \ln \frac{M^2_{K}}{\mu^2} +
3 M^2_{\eta} \ln \frac{M^2_{\eta}}{\mu^2} 
\right] \no \\
&& \mbox{} 
- \frac{8\sqrt{3}\;\varepsilon}{3F_0^2}L_5^r(\mu)(M_K^2-M_\pi^2)\no \\
&& \mbox{}
-\frac{\sqrt{3}\;\varepsilon}{4(4\pi)^2F_0^2}\left[ 
M^2_{\pi} \ln \frac{M^2_{\pi}}{\mu^2} - 
M^2_{\eta} \ln \frac{M^2_{\eta}}{\mu^2} 
-\frac{2}{3}(M_K^2-M_\pi^2)\left(\ln \frac{M^2_{K}}{\mu^2} + 1\right)
\right]\bigg\}.\label{FK1loop}\no \\
\eeqa
Performing the replacements $M_{\pi^\pm}\to M_{K^\pm}$ and 
$V_{ud}\to V_{us}$ in (7a) of \cite{Marciano/Sirlin}, we find that in 
the case of the kaon decay the constant $C_1$ corresponds to
\beq
C_1 = - 4\pi^2 E^r(M_{\rho}) - \frac{1}{2} 
+ \ln\frac{M_Z^2}{M_\rho^2}
+ \frac{Z}{4}\left[ 3 + 
\ln\frac{M_\pi^2}{M_\rho^2} + 2\ln\frac{M_K^2}{M_\rho^2}
\right].
\eeq

\section{Conclusions}
\label{sec: Conclusions}
\renewcommand{\theequation}{\arabic{section}.\arabic{equation}}
\setcounter{equation}{0}
We have developed the appropriate low-energy
effective theory for a complete treatment of isospin-violating
effects in semileptonic weak processes. The electromagnetic
interaction requires the inclusion of the photon and the
light lepton  fields as explicit dynamical degrees of freedom in the
chiral Lagrangian. At next-to-leading order, the list of local
terms given by Gasser and Leutwyler \cite{GL85} for the QCD part
and by Urech \cite{Urech} for the electromagnetic interaction of
the pseudoscalars has to be enlarged. This is, of course, a
consequence of the presence of virtual leptons in our extended
theory. Regarding pure lepton or photon bilinears as
``trivial'', five additional ``non-trivial'' terms of this type
are arising. Two of them will, however, not appear in realistic
physical processes. One may therefore conclude that the main
bulk of electromagnetic low-energy constants is already
contained in Urech's Lagrangian and the inclusion of virtual
leptons in chiral perturbation theory does not substantially aggravate
the problem of unknown parameters.

The continuation of the present work will follow two principal
lines. Firstly, we are now in the position to calculate the
electromagnetic corrections to semileptonic weak decays where
all constraints imposed by chiral symmetry are taken into
account. In spite of our large ignorance of the actual values 
of the electromagnetic low-energy couplings, it will often be
possible to relate the electromagnetic contributions to
different processes in this way. For specific combinations of
observables one might even find parameter-free predictions.
Simple examples of this kind have been given for the $P_{\ell
2}$ decays. In some fortunate cases simple order-of-magnitude
estimates for the electromagnetic couplings based on chiral 
dimensional analysis may even be sufficient. 
(See for instance \cite{NR95,ELM}.)  

The second major task for the near future is, of course, the
determination of the physical values of the coupling constants 
$K_i^r$ and $X_i^r$ in the standard model. In contrast to the
rather good information on the QCD effective couplings 
$L_1^r,\ldots, L_{10}^r$ that can be obtained in the standard
framework \cite{GL84,GL85}, only very
little is known so far in the electromagnetic sector. First attempts to
estimate some of the $K_i$ can be found in \cite{BB96,BB97,Moussallam}. As 
far as the constants $X_i$ are concerned, the recent 
analysis \cite{KPPdR99} of the counterterms contributing to the decay 
processes of light
neutral pseudoscalars into charged lepton pairs leads one to expect that 
reliable estimates for these constants can be achieved within a large-$N_c$
approach. Only with a more precise information about the
$K_i^r$ and $X_i^r$ will the electromagnetic part of the chiral
effective theory finally exhibit its full predictive power.

\section*{Acknowledgements}
We would like to thank G. Ecker for fruitful discussions and a careful reading 
of the manuscript.
P.T. acknowledges the kind hospitality of the FEN department, and expresses 
his thanks to the Swedish Research Council (NFR) for financial support.

\medskip\medskip
\newcounter{zaehler}
\renewcommand{\thesection}{\Alph{zaehler}}
\renewcommand{\theequation}{\Alph{zaehler}.\arabic{equation}}
\setcounter{zaehler}{1}
\setcounter{equation}{0}

\section*{Appendix}
\label{appdx}

The bosonic covariant derivative occurring in (\ref{L2E})
\beq
D_\rho = \partial_\rho + X_\rho
\eeq
has the index structure
\beq
X_\rho = - X_{\rho}^T = \left[ \ba{ll} (X_\rho)_{ij} & (X_\rho)_{i\nu} \\
(X_\rho)_{\mu j} & (X_\rho)_{\mu\nu} \ea \right], \label{X}
\eeq
where the matrix elements are given by
\beqa
(X_\rho)_{ij} &=& - \frac{1}{2} \langle \Gamma_\rho
[\lambda_i,\lambda_j]\rangle,  \no \\
(X_\rho)_{i\nu} &=& \frac{eF_0}{4} \delta_{\rho \nu}
\langle (\Q_R^{\rm em} - \Q_L^{\rm em})\lambda_i\rangle, \no \\
(X_\rho)_{\mu j} &=& - (X_\rho)_{j\mu}, \no \\
(X_\rho)_{\mu\nu} &=& 0.
\eeqa
The bosonic field $Y = Y^T$ has the same index structure as (\ref{X}) with
\beqa
Y_{ij} &=& s_{ij} - \frac{e^2 F_0^2}{4}
\langle \lambda_i(\Q_R^{\rm em} - \Q_L^{\rm em})\rangle
\langle \lambda_j(\Q_R^{\rm em} - \Q_L^{\rm em})\rangle, \no \\
Y_{i\nu} &=& - \frac{eF_0}{4} \langle \lambda_i \{ \nabla_\nu
(\Q_R^{\rm em} - \Q_L^{\rm em}) +
i[ u_\nu,\Q_R^{\rm em} + \Q_L^{\rm em}] \}\rangle, \no \\
Y_{\mu j} &=& Y_{j\mu}, \no \\
Y_{\mu\nu} &=& \frac{3e^2 F_0^2}{8} \delta_{\mu\nu}
\langle (\Q_R^{\rm em} - \Q_L^{\rm em})^2 \rangle.
\eeqa
The covariant derivative acting on the fermions,
\beq
\D_\rho = \partial_\rho + i f_\rho
\eeq
is determined by
\beq
f_\rho = \left[ \ba{ll} (f_\rho)_{ab} & (f_\rho)_{an} \\
(f_\rho)_{mb} & (f_\rho)_{mn} \ea \right] \label{f}
\eeq
with
\beqa
(f_\rho)_{ab} &=& - e A_\rho \delta_{ab}, \no \\
(f_\rho)_{an} &=& \frac{F_0^2}{2} \langle u_\rho \Q_L^{\rm w}\rangle
\delta_{an}, \no \\
(f_\rho)_{mb} &=& \frac{F_0^2}{2} \langle u_\rho \Q_L^{{\rm w}\dg}\rangle
\delta_{mb}, \no \\
(f_\rho)_{mn} &=& 0.
\eeqa
The fermion mass matrix $m$ shares the index structure with (\ref{f}).
The matrix elements read
\beq
m_{ab} = m_a \delta_{ab}, \qquad
m_{an} = m_{mb} = m_{mn} = 0.
\eeq
The fields $\alpha_\rho$, $\bar \alpha_\rho$, $\beta$, $\bar \beta$ are
fermionic quantities. $\alpha_\rho$ is given by
\beq
\alpha_\rho = \left[ \ba{ll} (\alpha_\rho)_{aj} & (\alpha_\rho)_{a\nu} \\
(\alpha_\rho)_{mj} & (\alpha_\rho)_{m\nu} \ea \right] 
\label{alpha}
\eeq
with
\beqa
(\alpha_\rho)_{aj} &=& - \frac{iF_0}{4} \langle [u_\rho, \Q_L^{\rm w}]
\lambda_j \rangle \nu_{aL} , \no \\
(\alpha_\rho)_{a\nu} &=& - \frac{e F_0^2}{4} \delta_{\rho \nu}
\langle \Q_L^{\rm w}(\Q_R^{\rm em} - \Q_L^{\rm em})\rangle \nu_{aL}
- e \delta_{\rho \nu} \ell_a, \no \\
(\alpha_\rho)_{mj} &=& - \frac{iF_0}{4} \langle [u_\rho, \Q_L^{{\rm w}\dg}]
\lambda_j \rangle \ell_{mL}, \no \\
(\alpha_\rho)_{m\nu} &=& - \frac{e F_0^2}{4} \delta_{\rho \nu}
\langle \Q_L^{{\rm w}\dg} (\Q_R^{\rm em} - \Q_L^{\rm em})\rangle
\ell_{mL}.
\eeqa
$\bar \alpha_\rho$ takes the form
\beq
\bar \alpha_\rho = \left[ \ba{ll} 
(\bar \alpha_\rho)_{ib} & (\bar \alpha_\rho)_{in} \\
(\bar \alpha_\rho)_{\mu b} & (\bar \alpha_\rho)_{\mu n} \ea \right]
\label{alphabar}
\eeq
with
\beqa
(\bar \alpha_\rho)_{ib} &=& - \frac{iF_0}{4}
\langle [u_\rho,\Q_L^{{\rm w}\dg}] \lambda_i\rangle \ol{\nu_{bL}}, \no \\
(\bar \alpha_\rho)_{in} &=& - \frac{iF_0}{4}
\langle [u_\rho,\Q_L^{\rm w}] \lambda_i\rangle \ol{\ell_{nL}}, \no \\
(\bar \alpha_\rho)_{\mu b} &=& - \frac{e F_0^2}{4} \delta_{\rho \mu}
\langle \Q_L^{{\rm w}\dg} (\Q_R^{\rm em} - \Q_L^{\rm em})\rangle
\ol{\nu_{bL}} - e \delta_{\rho\mu} \ol{\ell_b}, \no \\
(\bar \alpha_\rho)_{\mu n} &=& - \frac{e F_0^2}{4} \delta_{\rho \mu}
\langle \Q_L^{\rm w} (\Q_R^{\rm em} - \Q_L^{\rm em})\rangle
\ol{\ell_{nL}} .
\eeqa
Finally,
\beqa \label{beta}
\beta_{aj} &=& - \frac{F_0}{2} \langle \Q_L^{\rm w} \lambda_j\rangle
\nu_{aL} , \no \\
\beta_{mj} &=& - \frac{F_0}{2} \langle \Q_L^{{\rm w}\dg} \lambda_j\rangle
\ell_{mL} , \no \\
\beta_{a\nu} &=& \beta_{m\nu} = 0,
\eeqa
and
\beqa \label{betabar}
\bar\beta_{ib} &=& - \frac{F_0}{2} \langle \Q_L^{{\rm w}\dg} \lambda_i\rangle
\ol{\nu_{bL}} , \no \\
\bar\beta_{in} &=& - \frac{F_0}{2} \langle \Q_L^{\rm w} \lambda_i\rangle
\ol{\ell_{nL}} , \no \\
\bar \beta_{\mu b} &=& \bar \beta_{\mu n} = 0.
\eeqa


\end{document}